\documentclass{cimento}
\usepackage{graphicx}        
\usepackage{color}           
\usepackage{url}             

\title{Christopher Clavius astronomer and mathematician}

\author{C.~Sigismondi\from{ins:i}\from{ins:a}\from{ins:u}\from{ins:e}}

   \instlist{\inst{ins:i} ICRA, International Center for Relativistic Astrophysics, and Sapienza University of Rome, P.le Aldo Moro 5 00185, Roma (Italy)
                    
             \inst{ins:a} Universit\'e de Nice-Sophia Antipolis, Laboratoire H. Fizeau, Parc Valrose, 06108, Nice (France)
                     
             \inst{ins:u} Istituto Ricerche Solari di Locarno, via Patocchi 57 6605 Locarno Monti, (Switzerland)
                    
             \inst{ins:e} GPA, Observat\'{o}rio Nacional, Rua General Cristino 77 20921-400, Rio  de Janeiro (Brazil)
                  }

\PACSes{\PACSit{01.65.+g}{History of science}}

\begin{document}

\maketitle

\begin{abstract}

The Jesuit scientist Christopher Clavius (1538-1612) has been the most influential teacher of the renaissance. His contributions to algebra, geometry, astronomy and cartography are enormous. He paved the way, with his texts and his teaching for 40 years in the the Collegio Romano, to the development of these sciences and their fruitful spread all around the World, along the commercial paths of Portugal, which become also the missionary paths for the Jesuits.
The books of Clavius were translated into Chinese, by one of his students Matteo Ricci "Li Madou" (1562-1610), and his influence for the development of science in China was crucial. The Jesuits become skilled astronomers, cartographers and mathematicians thanks to the example and the impulse given by Clavius.
This success was possible also thanks to the contribution of Clavius in the definition of the {\it ratio studiorum}, the program of studies, in the Jesuit colleges, so influential for the whole history of modern Europe and all western World.

\end{abstract}

\section{Introduction}

The books of Clavius traveled all around the World with Jesuits in the same years of their publication.
On the marine roads toward China, the Italian Jesuit Matteo Ricci from Macerata, brought the Euclides' Geometry edited by Clavius, and translated it into Chinese. Also the Clavius name become the Chinese "KeLaWeiWuSi" In the territories of Brazil from 1750 to 1759 Ignazio Szentmartonyi, Croatian of Kotiri, was measuring the limits, for the Portuguese governement.\cite{leite}
Clavius wrote also a booklet {\it Geometria Practica} with simplified concepts for China,\cite{practica} upon request of Matteo Ricci.
At Collegio Romano in Rome, Clavius taught Mathematics for more than 40 years, from 1564, raising a school of mathematicians astronomers and cartographers spread all over the World.
For this reason he was defined as the most influential teacher of Renaissance by George Sarton.\cite{schweitzer}
His books on Arithmetic, Geometry, Algebra, Harmonics and Astronomy were used in all the
European Jesuit schools.
Clavius was the mathematics instructor of Catholic Europe as well as much of Protestant
Europe.

\begin{figure}
\centerline{\includegraphics[width=1\textwidth,clip=]{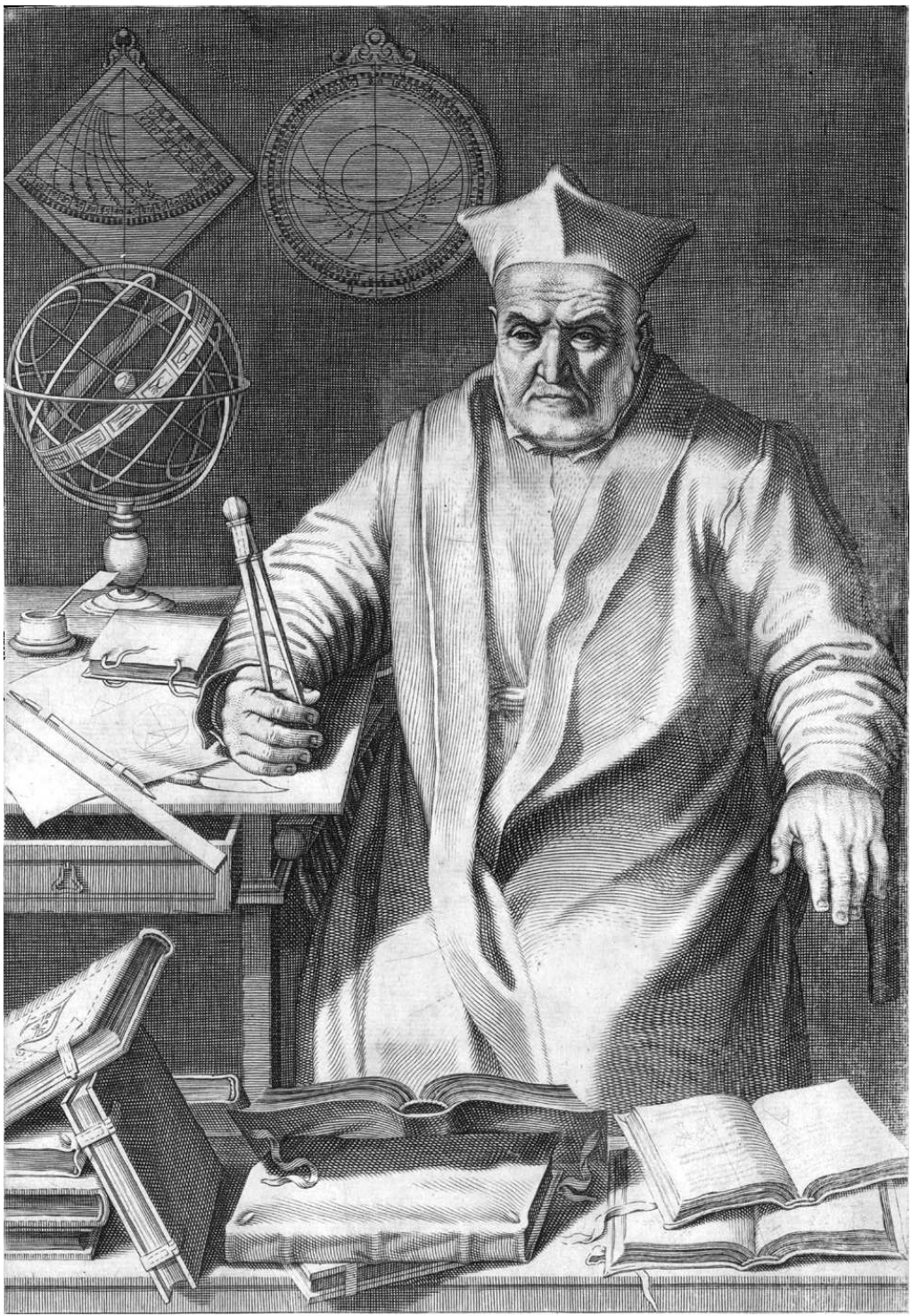}}
\caption{Christopher Clavius (1538-1612) born in Bamberg (Germany), studied in Coimbra (Portugal)and taught Mathematics, Geometry, Astronomy and Cartography in the Collegio Romano, in Rome (Italy) all his life.}
\label{Fig. 1}
\end{figure}

\section{Algebra}
Christopher Clavius published the first edition of his algebra textbook in 1608, in
Rome, by Zannetti.\cite{algebra}
Leibnitz wrote a letter to Bernoulli in 1703,
indicating that he learned algebra from Clavius' book. {\it As a child, I had studied the
elementary algebra of one Lancius, and later that of Clavius …}.

The task of algebra, solving equations, had to realize four stages: firstly, putting the verbally
formulated, word problem, into an equation; next to reduce the equation, that is to put homogeneous
terms on the same side; thirdly to divide the highest power of the equation by its coefficient to have the
unknown not affected by coefficients other than 1; and lastly, if necessary, to extract the respective root
in order to get the value for the unknown.

This is surprisingly unsophisticated. Basically, it corresponds to what Al-Khwarizmi had already given as
the rule: {\it al-jabr} and {\it al-muqabala} are just these reduction procedures. Clavius, hence, does not present a
theoretical program for algebra: it is rather a set of practical rules. In fact, there do not follow
systematizations of types of equations and standard procedures for their solution.
The book continues to present and discuss examples of problems. In fact, the bulk of the remainder of
the volume, with its 383 pages, consists of four chapters with an immense number of problems and their
solutions. Characteristically, the titles of these three chapters begin by {\it Aenigmata}, i.e. riddles. This
recalls the style and contents of Arab treatises and Italian treatises, from Fibonacci on, in particular of
the {\it trattati d'abbaco}. Problems used to be given there in a recreational manner.\cite{schubring}

Clavius calculated and published a table of sines and cosines with 7 decimal places in his book
{\it Astrolabium}. For each value of degrees and minutes, the table gives the value of the sine
and cosine with the precision of one part in ten million. Unlike other tables then in
circulation, he was very careful to avoid errors. He showed how to interpolate to get precise
values for fractions of seconds of arc. This significantly shortened computations in astronomy
and spherical geometry.

Clavius was the first to use a decimal point, some twenty years before it became common;
the first to use parentheses to collect terms;
and the first to use the plus and minus signs + and - in Italy.

Likewise, he used no sign for equality and expressed it verbally,
too. For instance, for establishing an equation, he put: {\it aequatio sit inter 5x + 7 \& 1x}.

\begin{figure}
\centerline{\includegraphics[width=1\textwidth,clip=]{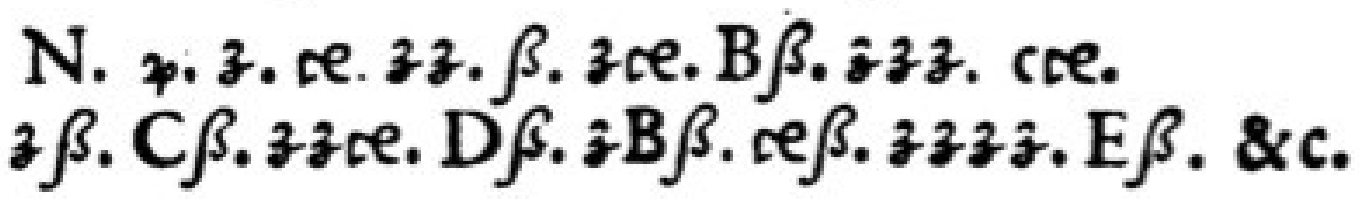}}
\caption{Here the successive symbols represent the
powers of x, that is, $1$, $x$, $x^2$, $x^3$, and so on. As
the {\it etc.} (\&c) at the end of the list shows,
the powers could proceed without a limit.
Then (see the image on the right), he
explained the meaning of these signs,
beginning from N as the unity, and continuing
with the sign for {\it res}. In fact, the
strange twirl, presenting $x$, is a ligature of
{\it re(s)}. The sign for $x^2$ represents a {\it z} in
some script font, and is derived from census,
the Latin expression used for square.
Eventually, he gives a list of the first twelve
signs, i.e. of the first 12 powers of the
variable.
It is easy to see that the terminology for these algebraic entities is derived from geometric notions:
square, cube, and building higher powers from compound forms. On the other hand, this geometric
language did not keep him from constructing an unlimited series of powers. Actually, the letters D, E,
etc. used for higher powers, were no longer taken from a geometric context.}
\label{Fig. 2}
\end{figure}

\section{Geometry and Cartography}
Clavius was called {\it the Euclid of the 16th century} by contemporary scientists.
The Commentary on Euclid's Elements by Clavius first edited in 1574\cite{euclide}
was not merely Euclid's Elements of Geometry, but a careful commentary;
a compendium of all the geometry known at the time.
It contained many original theorems as well as some axioms Euclid missed.
Six other editions along his life appeared: "the large number of editions which were needed to satisfy the demand for his Euclid
shows the high reputation which the work achieved" (Moritz Cantor).\cite{schweitzer}
Clavius' Euclid became the standard geometry textbook for the 17th century.

\begin{figure}
\centerline{\includegraphics[width=0.2\textwidth,clip=]{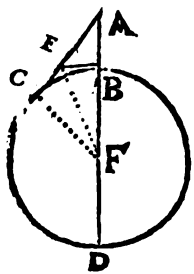}}
\caption{In Geometria Practica Clavius stated and showed how to solve dozens
of problems, such as: Problem 20 (p. 214): To calculate the size of the Earth (its radius BF=DF=CF, in the figure
above) from observations at the peak (A) of a mountain whose height (AB) is known.}
\label{Fig. 3}
\end{figure}

\section{Astronomy: Commentarius to the Sphera and the Gregorian Calendar}

In the same way he did with Geometry he wrote and extended Commentary\cite{sfera} to the small textbook of John Holywood (1256):
the {\it Sphaera}. The first edition appeared in 1570 in Venice.
The original Sphaera was merely a list of astronomical and geographical topics, with very incomplete treatments.
And Galileo Galilei reading the Commentarius made by Clavius could say that there was a great qualitative 
difference between it and the {\it Sphaera}.\cite{sigismondi} 
The Commentarius was finally the first real modern textbook of Astronomy:
almost 500 pages, with complete geometrical and mathematical treatments of all topics. 

Clavius showed always an acute critical faculty. There is no difficulty that he attempted to evade...neither any established authority
(like Ptolemy or Kepler) which cannot be discussed.
The case of the annular eclipse of 9 April 1567, observed from Rome and reported in the 1581 edition of the Commentarius, is remarkable.
According to Ptolemy's {\it Almagest} an annular eclipse was impossible, but Clavius assessed his observation as of an annular eclipse,
reporting this phenomenon for the first time in a scientific publication.
Kepler believed that the Moon could have an atmosphere, for this reason send two persons to ask Clavius about the nature of that eclipse.
One of them was Johannes Remus. Clavius always said that the 1567 eclipse was annular. 
Nowadays we know that it was hybrid, and a long debate was held between J. Eddy and Stephenson, Jones and Morrison on the possible change in 
solar diameter from 1567, at the end of Sp\"{o}rer Grand Minimum, and today.\cite{eddy,morrison}

\begin{figure}
\centerline{\includegraphics[width=1\textwidth,clip=]{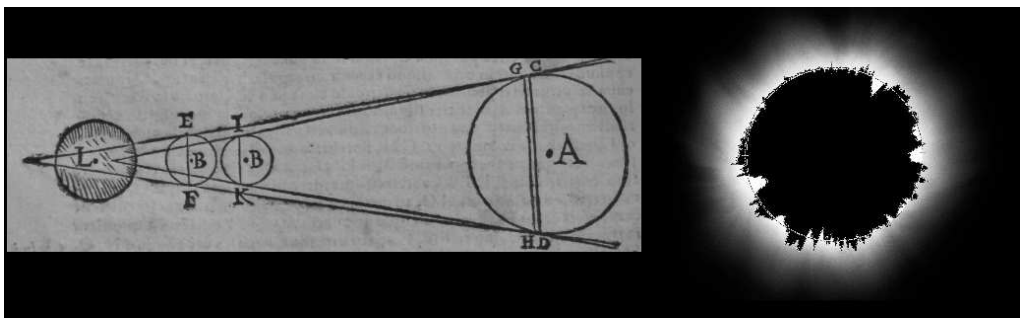}}
\caption{The eclipse of 1567 in a photocomposition. The appearance of a ring it is due to the solar mesosphere, where thousands of tiny emission light yield an 
appearance of white light.\protect\cite{bazin}}
\label{Fig. 4}
\end{figure}

Another monumental opera was the apology of Gregorian Calendar,\cite{calendar} where he exemplified all consequences of 
the reformation of the calendar issued in 1582.
Clavius wrote also the {\it Astrolabium},\cite{astrolabium} and fundamental texts on sundials.\cite{gnomonices,horo}

\begin{figure}
\centerline{\includegraphics[width=1\textwidth,clip=]{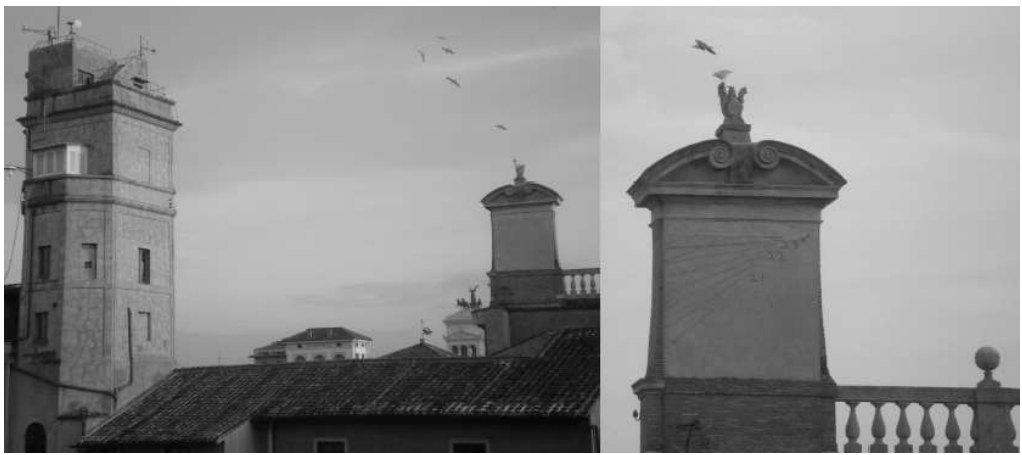}}
\caption{The horologium of Clavius at Collegio Romano. It works with Italic hours, i.e. sunset corresponds to the $24^{th}$ hour.}
\label{Fig. 5}
\end{figure}

To him is dedicated a crater on the Moon, in the Selenography made by Giovanni Battista Riccioli 
and Francesco Grimaldi in the {\it Almagestum Novum} of 1655.
The crater is the bigger in the lunar sector dedicated to Ptolemaic astronomers, and Clavius was indeed the greatest among them.



\bibliography{Clavius}

\begin{thebibliography}{1}
\expandafter\ifx\csname url\endcsname\relax\def\url#1{\texttt{#1}}\fi
\expandafter\ifx\csname urlprefix\endcsname\relax\def\urlprefix{URL }\fi

\bibitem{leite}
\NAME{{Leite} S.}, \TITLE{{Historia da Companhia de Jesus no Brasil}},
 Lisboa - Rio de Janeiro, 10 volumes, 1938-1950.

\bibitem{practica}
\NAME{{Clavius} C.}, \TITLE{{Geometria Practica}},
Rome, 1604.

\bibitem{schweitzer}
\NAME{{Schweitzer} P.}, \TITLE{{Contributions of Christopher Clavius S.J. to Mathematics}},
  \IN{Gerbertus}{2}{2012} in press.

\bibitem{algebra}
\NAME{{Clavius} C.}, \TITLE{{Algebra}},
Geneva, 1609.

\bibitem{schubring}
\NAME{{Schubring} G.}, \TITLE{{The Book "Algebra" of Christopher Clavius S.J.}},
  \IN{Gerbertus}{2}{2012} in press.

\bibitem{euclide}
\NAME{{Clavius} C.}, \TITLE{{Euclidis Elementorum Libri XV}},
 Venice, 1574.

\bibitem{sfera}
\NAME{{Clavius} C.}, \TITLE{{In sphaeram Ioannis de Sacro Bosco Commentarius}},
 Venice, 1581.

\bibitem{sigismondi}
\NAME{{Sigismondi} C.}, \TITLE{{La Sfera da Gerberto al Sacrobosco}},
  APRA Rome, 2008.
  
\bibitem{bazin}
\NAME{Sigismondi, C., A. Raponi, C. Bazin and R. Nugent},  \IN{arXiv:1106.2197} (2011). 
  
\bibitem{eddy} 
\NAME{Eddy, J.,  A. A. Boornazian, C. Clavius, I. I. Shapiro, L. V. Morrison and S. Sofia}, {\it Sky
\& Telescope} {\bf 60}, 10 (1980).

\bibitem{morrison} 
 \NAME{Stephenson, F. R., J. E. Jones and L. V. Morrison}, 
 \IN{Astron. \& Astrophys.} {\bf 322} 347 (1997).  
 
\bibitem{calendar}
\NAME{{Clavius} C.}, \TITLE{{Romani calendarii a Gregorio XIII restituti explicatio}},
Rome, 1603.
 
\bibitem{astrolabium}
\NAME{{Clavius} C.}, \TITLE{{Astrolabium}},
Rome, 1593.

\bibitem{gnomonices}
\NAME{{Clavius} C.}, \TITLE{{Gnomonices Libri Octo}},
Rome, 1581.
 
\bibitem{horo}
\NAME{{Clavius} C.}, \TITLE{{Horologiorum nova descriptio}},
Rome, 1599.
 

\end{thebibliography}
   
\bibliographystyle{varenna}

\end{document}